\documentclass[]{spiejour}  %>>> use for US letter paper
\usepackage[]{graphicx}

\title{Optical properties of cosmic dust analogs: A review}

\author{Thomas Henning\supscr{a} and Harald Mutschke\supscr{b}}

\affiliation{\supscr{a}Max-Planck-Institut f\"ur Astronomie, K\"onigstuhl 17,
D-69117 Heidelberg\\
\linkable{henning@mpia-hd.mpg.de} \\
\supscr{b}Astrophysical Institute and University Observatory (AIU),
Friedrich Schiller University, Schillerg\"a\ss{}chen 3, D-07745
Jena, Germany\\
\linkable{mutschke@astro.uni-jena.de}
}

\begin{document}
  \maketitle

%%%%%%%%%%%%%%%%%%%%%%%%%%%%%%%%%%%%%%%%%%%%%%%%%%%%%%%%%%%%%
\begin{abstract}
Nanometer- and micrometer-sized solid particles play an important role in
the evolutionary cycle of stars and interstellar matter. The optical
properties of cosmic grains determine the interaction of the radiation field
with the solids, thereby regulating the temperature structure and
spectral appearance of dusty regions. Radiation pressure on dust grains 
and their collisions with the gas atoms and molecules can drive powerful 
winds. The analysis of observed spectral features, especially in the 
infrared wavelength range, provides important information on grain size, 
composition and structure as well as temperature and spatial distribution 
of the material. 

The relevant optical data for interstellar, circumstellar, and protoplanetary
grains can be obtained by measurements on cosmic dust analogs in 
the laboratory or can be calculated from grain models based on optical 
constants. Both approaches have made progress in the last years, triggered 
by the need to interpret increasingly detailed high-quality astronomical 
observations. The statistical theoretical approach, spectroscopic 
experiments at variable temperature and absorption spectroscopy of aerosol 
particulates play an important role for the successful application of the data 
in dust astrophysics.

\end{abstract}

%>>>> Include a list of up to six keywords after the abstract

%%%%%%%%%%%%%%%%%%%%%%%%%%%%%%%%%%%%%%%%%%%%%%%%%%%%%%%%%%%%%
\section{INTRODUCTION}
\label{sect:intro}  % \label{} allows reference to this section
Dust grains exist in a wide variety of cosmic environments ranging
from the diffuse interstellar medium and molecular clouds in our galaxy
to circumstellar shells and disks around young stellar objects, dusty 
outflows from evolved stars, atmospheres of brown dwarfs, the dust tori 
of active galactic nuclei and starburst galaxies, and high-redshift quasars. 
The particles are especially important during the birth process of stars 
and as precursor material for the formation of planets. In the late stages 
of stellar evolution they condense in molecular outflows and are important 
for the mass loss process and the spectral appearance of AGB and post-AGB 
stars.

The field of cosmic-dust research has been the subject of intense research 
over the last decades. The international conferences ``Astrophysics of Dust'' 
in Estes Park, Colorado (2003) and ``Cosmic Dust - Near and Far'' in 
Heidelberg, Germany (2008) both reflected the rapid development of the field. 
Astronomical dust research is of course triggered by the progress in 
astronomical observations, mainly by those in the infrared range, because 
of the opportunities to (1) observe dusty objects optically thick in the 
visible and (2) to analyse the thermal emission of dust by infrared 
spectroscopy. The ISO and Spitzer space telescopes have played a major 
role in this respect, and current and future projects such as the Herschel 
mission and JWST will continue to ensure progress in this field.

The description of the interaction of radiation fields with a system of
solid particles always requires knowledge of their absorption and
scattering cross sections which depend on their chemical composition,
solid-state structure and morphology (see Table~\ref{tbl-1}).
Any continuum  radiative transfer code for the calculation of
intensity/polarization maps and spectral energy distributions needs
the specification of these properties. In addition, the transfer of
momentum from a radiation field to a grain is determined by
the extinction and scattering  efficiency and the average cosine of the
scattering angle.

\begin{table}
\caption{Chemical and structural properties of solid particles and
the description of optical properties on different length scales } \label{tbl-1}
\begin{center}
\begin{tabular}{lll}
Length scale &  Structural properties &  Description\\
&&\\
\hline
&&\\
``Atomic''          & Chemical composition and bonding  &  Dielectric function $\epsilon(\lambda)$\\
                    & Crystal structure                 &  Complex refractive index $m(\lambda$)\\
                    & Defects and impurities            &        \\
&&\\
``Mesoscopic''     & Inhomogeneities                   &  Theory of effective media\\
                    & Porosity                          &  $\epsilon_{\rm eff}(\lambda)$,\\
                    & Mantles                           &  Core-mantle description\\
                    & Surface states                    &  \\
&&\\
``Macroscopic'',    & Size distribution                 &  Mie theory, T-matrix method,\\
 Morphology         & Grain shapes                      &  Separation of variables method,\\
                    & Agglomeration                     &  Discrete dipole approximation\\
                    & Coalescence        &  $C_{\rm ext}(\lambda), C_{\rm abs}(\lambda), C_{\rm sca}(\lambda)$\\
\end{tabular}
\end{center}
\end{table}

In this review we will deal with the progress in theoretical and 
experimental work aimed to providing data and tools for the interpretation 
of spectroscopic dust-related observations, and for modeling of dusty 
astrophysical objects. Current models go significantly beyond the
assumption of spherical grains characterized by artificially constructed optical 
constants, as has been the prefered approach for many years. Effects of grain
morphology, temperature, and composition have been studied experimentally,
and complex light-scattering models have been used to explore the dependence
of scattering and absorption cross sections on the structure of grains
on all length scales (see Table~\ref{tbl-1}), including porosity,
inhomogeneity, and agglomeration. Such grain models have been used in 
sophisticated multi-dimensional radiative transfer codes. Among the 
experimental data, which can be included as inputs into the computations, 
are also measurements of the scattering matrices of cosmic dust analogs 
which have been performed in Amsterdam for many years and have been made 
available through the Amsterdam Light Scattering Database 
(http://www.astro.uva.nl/scatter). Although this review unfortunately 
cannot cover this and some other experimental fields devoted to interpret 
observational data of cosmic dust, all general considerations presented 
in the next sections are valid for them as well, especially those on 
the relation between calculations and measurements. 

It is beyond the scope of this paper to discuss extensively the
astronomical dust models and optical data derived from observations. 
For a description of the various ``classical'' dust models, we refer 
to the contributions in Ref. \nocite{alti89}1. 
More recent models have been developed by Mathis \cite{mathis96},
Li \& Greenberg \cite{ligb97}, Zubko \textit{et al.} \cite{zubko04}, 
Draine \& Li \cite{drli07} and can also be found in the proceedings of 
the two dust conferences \cite{AoD03,DNF08}.

\section{Two ways for predicting optical data of cosmic dust particles}
\label{sect:data}

\begin{figure}
\centering
\includegraphics[height=9cm]{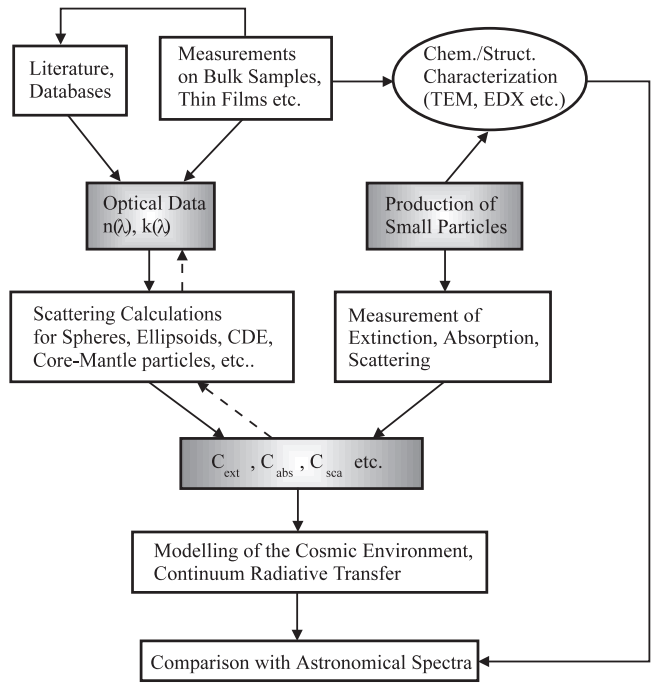}
\caption{Two ways for determining optical properties of cosmic dust analogs.}
\label{fig-1}
\end{figure}

Our knowledge about the nature of cosmic dust grains mainly
comes from spectroscopic observations over the infrared, optical,
and UV wavelength ranges. Infrared spectroscopy provides
information about vibrational modes of functional groups and
crystal lattices, whereas the optical and UV range of the galactic
extinction curve is mainly dominated by a strong UV resonance at
217 nm, usually attributed to an electronic transition in
carbonaceous grains or polyaromatic hydrocarbons (PAHs). 
In the infrared, features related to the abundant silicates 
are frequently observed, especially the bands around 10 and 18~$\mu$m 
attributed to Si-O stretching and bending vibrations. 

Other constraints for the chemical and physical structure of 
grains are provided by abundance considerations, by our understanding 
of the temperature and density range, as well as by the presence of 
shocks and winds in interstellar and circumstellar media (see, e.g., 
Refs. \nocite{dohe95,vdishoeck04}8 and 9). In oxygen-rich environments 
we can expect the presence of oxide materials (mainly silicates) and
oxygen-containing ices (H$_2$O, CO$_2$). Carbon-rich environments 
contain mainly carbonceous solids, carbides, and sulfides. 
Here, we will only deal with refractory materials and not 
with molecular ices \cite{schutte98,schutte99} and larger molecules 
such as PAHs \cite{hesal98,salama99}.

After the discovery of abundant crystalline minerals in circumstellar disks
(e.g., Refs. \nocite{jaeger98a,waters10}14 and 15), and the increasing efforts to study the 
detailed mineralogy and morphology of the dust in these environments from 
its infrared emission spectra, both theorists working on light-scattering 
theory and experimentalists in astrophysical laboratories have focused their
work on predicting the spectroscopic properties of cosmic dust particles
more accurately. In contrast to the amorphous silicates in the diffuse 
interstellar medium, crystalline silicates are frequently observed in 
protoplanetary disks, which indicates their in-situ formation by thermal 
annealing or shocks (for a review see Ref. \nocite{heme09}16). An interesting result 
of astronomical observations of protoplanetary and circumstellar dust 
is the fact that the crystalline silicates are mostly magnesium-rich 
\cite{bouwman01,bouwman08,juhasz10}. 

Extracting reliable dust properties from the thermal emission spectra of 
a mixture of dust particles requires precise knowledge of the optical 
properties for all possible constituents under various conditions. 
Crystalline silicates as the most important dust minerals, e.g., have 
very characteristic infrared spectra with many bands at wavelengths 
between 8~$\mu$m and the submillimeter range. The details of band 
positions and band profiles, however, depend on the chemical composition, 
the grain morphology, and the grain temperature.
This means that morphological information, e.g., can only been extracted from
the spectra if the influence of the other parameters is sufficiently well
understood and can be disentangled. 

The grains condensed in circumstellar environments and modified by 
interstellar processes and even re-formed in the interstellar medium 
may be quite chaotic in composition and structure. Table~\ref{tbl-1} 
summarizes some of the chemical and structural properties which may 
characterize the particles on different length scales. The theoretical 
description of the optical properties of such particles can be very 
difficult. Even the ``optical constants'' (the complex refractive 
index $m=n+ik$ or the dielectric function 
$\epsilon=\epsilon_1 + i \epsilon_2$) of the materials constituting the
particle may be poorly known.

These facts require studies which have to be done in the laboratory,
via synthesis and optical spectroscopy of materials representing
important aspects of the cosmic solid matter \cite{colangeli03,jaeger09}. 
The investigations of  these ``cosmic dust analogs'' do not necessarily 
aim at exactly reproducing cosmic dust. In our terminology dust analogs 
comprise also bulk matter or thin films prepared for the derivation of 
optical material constants. In addition, we also consider simple 
``model'' particles (e.g., silica monospheres) used for the investigation 
of special aspects of particle optics (e.g., agglomeration effects).

Figure~\ref{fig-1} illustrates the various aspects of this approach. 
Two principal possibilities exist for the derivation of the optical 
properties of particles:

\begin{enumerate}
\item Calculation of cross sections from optical material data which
may have been measured or taken from a data base. In most cases, the
data will originate from a bulk or thin-film material since measurements
on these geometrically simple systems are much easier and much more work
has been done on them. Therefore one has the problem that the material
may not be exactly the one one would like to have. Special material
structures may only exist in the form of small grains (e.g., silicon
nanoparticles, carbon onions). Additionally, theory has some problems 
with the calculation of spectra for complicated grain shapes in
strong absorption bands. This will be discussed further in Sec.~4.
\item Direct measurement of the optical properties on particle samples.
This certainly is the preferred but also experimentally much more
demanding possibility. One has to produce particles of the desired
composition and structure and one has to store them in an environment
as similar to the cosmic one as possible. This is sometimes problematic 
because conventional preparation of particulate samples for spectroscopy 
(i) uses matrices in which the particles are embedded and 
(ii) is not able to prevent agglomeration of the primary particles to 
larger clusters.
\end{enumerate}

The next section will treat the basics of the first (computational) way,
the light scattering theory, which  also underlies the experimental
way. Section 4 will deal with measurements of the material optical data
from bulk and thin film samples as well as of extinction data of particulate
samples.

%%%%%%%%%%%%%%%%%%%%%%%%%%%%%%%%%%%%%%%%%%%%%%%%%%%%%%%%%%%%%
\section{Astronomical features and developments in light scattering theory} \label{sect:theory}

The optical properties of small particles can considerably deviate
from those of bulk materials because of the occurrence of surface
modes. The structure of the interface of small particles with the
surrounding medium usually has strong effects on their optical
properties. In addition, the shape and internal structure of the
particles are important parameters in determining the position and
shape of the observed astronomical features. Furthermore, the
optical depth and the temperature structure of the dusty
astronomical media are decisive for the occurrence of the
solid-state bands and will determine if they are observed in
emission, absorption, or self-absorption. Protostars, deeply
embedded in their parental molecular clouds, show absorption
features due to molecular ices (e.g., H$_2$O, CO, CO$_2$ and
NH$_3$) and refractory silicates. In contrast, the spectra of
protoplanetary disks around young stars are mostly characterized
by a variety of silicate emission bands, produced in the optically
thin disk atmosphere on top of the optically thick disk region.

For micron-sized grains classical electrodynamic theory can be
applied using bulk optical constants. Textbooks which treat the
classical electrodynamics of absorption and scattering of light by
small particles are the publications by van de Hulst \cite{hulst57},
Kerker \cite{kerker69}, Bohren \& Huffman \cite{bohren83}, Kokhanovsky 
\cite{kokh99}, Mishchenko, Hovenier \& Travis \cite{mishch00}, 
and Mishchenko, Travis \& Lacis \cite{mishch02} .

The qualitative features of the absorption and scattering of light strongly
depend on the ratio between the wavelength of the incident light $\lambda$ and
the typical radius of the particle $a$. We can distinguish three regions:

\begin{enumerate}
\item $\lambda$ $\ll$ $a$: This is the limit of geometrical
optics, where the propagation of light is described by rays which
are reflected and refracted at the surface of the scatterer
according to Snell's law and the Fresnel formulae. For absorbing
materials, light can penetrate only within the skin depth.
Scattering, therefore, is mainly a surface effect and the
extinction cross-section per volume goes roughly as $1/a$. 
\item $\lambda$ $\sim$ $a$: This is the regime of wave optics, 
where both the angular and the wavelength dependence of the cross
sections are dominated by interferences and resonances. For
spherical particles, these can be calculated by the Lorenz-Mie 
theory \cite{mie08}.
\item $\lambda$ $\gg$ $a$: This is the Rayleigh limit. If
furthermore we have $\lambda$ $\gg$ $|m| a$, where $m$ is the
(complex) refractive index of the particle, we are in the 
quasi-static limit. Then, both the incident and the internal 
field can be regarded as static fields. In this regime,
phase shifts over the particle size are negligible. This implies
that it is generally sufficient to consider only the dipolar
electric mode.
\end{enumerate}

The interaction of infrared radiation with sub-micron sized cosmic
grains can generally be considered as a good example of the
quasi-static case. However, particles with high imaginary parts of
the refractive index (metals, semiconductors) may easily violate
this condition even at infrared wavelengths. The same is true for
laboratory measurements where it is not always guaranteed that the
particles fulfill the condition for the quasi-static case. 

\begin{figure}
\centering
\includegraphics[height=9cm]{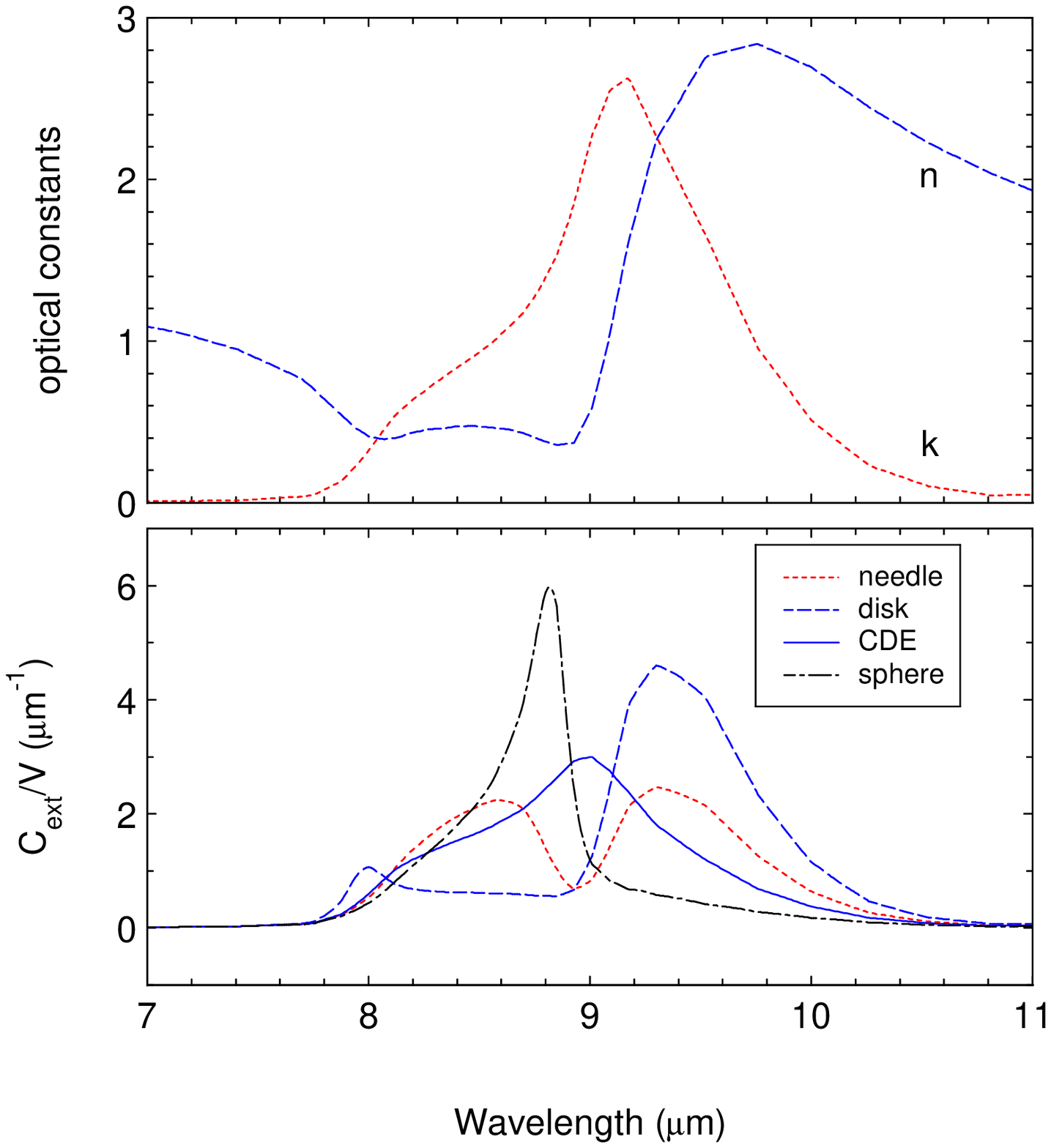}
\caption{Shape effects for SiO$_2$ particles in the quasi-static limit.
CDE refers to a continuous distribution of ellipsoidal shapes (Eq. 9).} 
\label{fig-2}
\end{figure}

For the sake of simplicity and physical insight, we will deal in the following
only with the quasi-static case. The extinction cross section $C_{\rm ext}$ is
given by:
\begin{equation}
C_{\rm ext} = C_{\rm abs } + C_{\rm sca},
\end{equation}
where $C_{\rm abs}$ and  $C_{\rm sca}$ are the absorption and scattering
cross sections.  According to Rayleigh's law $C_{\rm sca}$ $\propto$ $k^4$
with the wavenumber $k=2\pi/\lambda$. For $\lambda$~$\gg$~$a$ this means in
practice that the extinction cross
section equals the absorption cross section. The cross section for the
radiation pressure can be computed via the formula
\begin{equation}
C_{\rm pre} = C_{\rm ext} - C_{\rm sca} \langle {\rm cos}\theta \rangle,
\end{equation}
with $\langle {\rm cos}\theta \rangle$ the mean of the cosine of the scattering
angle. Following the conventions in \cite{bohren83},
the expression for the extinction cross section is given by:
\begin{equation}
C_{\rm ext}= k {\rm Im}(\alpha),
\end{equation}
with  $\alpha$ being the polarizability. This quantity depends on the complex
dielectric function, $\epsilon$ = $\epsilon_1$ + i $\epsilon_2$, of the particle
material and on the dielectric function of the embedding medium,
$\epsilon_{\rm m}$, which is in most cases a real wavelength-dependent number.
In addition, the polarizability  is determined by the shape of the particle.

For ellipsoids the polarizability, $\alpha_{\rm i}$, in an electric field parallel
to one of its principal axes is given by
\begin{equation}
\alpha_{\rm i} = V \frac{\epsilon - \epsilon_{\rm m}}{\epsilon_m + L_{\rm i} (\epsilon - \epsilon_{\rm m})},
\end{equation}
where $L_{\rm i}$ are geometrical factors and V is the volume of the ellipsoid.
The relation $L_1$+$L_2$+$L_3$=1
implies that only two of these factors are independent. In the case of prolate
spheroids with equal minor axes of length b=c and with eccentricity
$e = \sqrt{1-c^2/a^2}$ (a the length of the major axis) the geometrical
factors are given by
\begin{equation}
L_1 = \frac{1-e^2}{e}\left(-1 + \frac{1}{2e} {\rm ln} \frac{1+e}{1-e}\right)
\end{equation}
and $L_2$=$L_3$=$(1-L_1)/2$. In the other case of oblate spheroids we have
$a$=$b$ and $L_1$=$L_2$=$(1-L_3)/2$ with
\begin{equation}
L_1 = \frac{\sqrt{e^{-2}-1}}{2e^2}\left(\frac{\pi}{2} - {\rm tan}^{-1}(\sqrt{e^{-2}-1})\right) - \frac{e^{-2}-1}{2}
\end{equation}

The simplest case of light scattering in the quasi-static limit is that for a
sphere where $a=b=c$ and $L_{\rm i}$ = 1/3. This gives $\alpha_i$=$\alpha$
\begin{equation}
\alpha =  3 V \frac{\epsilon - \epsilon_{\rm m}}{\epsilon + 2 \epsilon_{\rm m}}.
\end{equation}
These equations demonstrate that resonances for particles surrounded by a
non-aborbing medium occur close to  wavelengths
where the imaginary part of the dielectric function is close
to zero and the real part fulfills the condition
\begin{equation}
\epsilon_1 = \epsilon_m (1-L_{\rm i}^{-1}).
\end{equation}
This immediately implies that the wavelengths of resonances depend
on the shape of the particles. The resonances can only occur in
regions where $\epsilon_1$ is negative (for a sphere 
$\epsilon_1 = - 2 \epsilon_{\rm m}$). Examples of astronomically relevant
materials which have negative values of $\epsilon_1$ are SiC 
\cite{mutschke99} and SiO$_2$ \cite{hemu97} in the infrared and 
graphite in the UV \cite{TaftPhil65}. For metal particles 
resonances even occur at wavelengths where the bulk material does 
not show any absorption band. In contrast, the lattice features 
of amorphous silicates do not show this behaviour. 

In the case of lattice modes, the resonances are located between 
the transverse and logitudinal optical phonon frequencies. As an 
example, we show shape effects for SiO$_2$ particles in 
Fig.~\ref{fig-2}. The equations also show that the positions of
resonances depend on the surrounding medium - a fact which is
often considered in astronomy in the case of dust models with
core-mantle grains. For particles of arbitrary shape there are no
analytical solutions of the light scattering problem (even in the
quasi-static limit) available. Numerical models frequently used
for non-spherical particles are the separation of variables
method, the T-matrix method and the discrete dipole (or multipole)
approximation  (see, e.g., Ref. \nocite{voshch00}32). The treatment 
of anisotropic materials gets more complicated, but is still 
possible \cite{michel99}.

In reality, we always expect a distribution of particle shapes. The
difficulty both in astronomical applications and in laboratory experiments
is to evaluate this shape distribution. A simple model for a distribution 
of particle shapes is the continuous distribution of ellipsoids (CDE), 
which by averaging Eq. (4) over all L$_i$ leads to
\begin{equation}
\alpha_{av} =  V \frac{2 \epsilon}{\epsilon - 1} {\rm Log} {\epsilon},
\end{equation}
where Log $\epsilon$ denotes the principal value of the logarithm of the
complex number $\epsilon$ (see Bohren \& Huffman 1983). 
The CDE assumes equal probability for the presence of every shape 
and averages over all orientations. This model can be used to estimate 
in an approximate way how important shape effects for a special material 
could be. 

Recently, Min \textit{et al.} \cite{min06a} have shown that an arbitrary grain 
shape can be described by a distribution of geometrical factors P(L), 
which can be derived from a spatial discretization of the particle shape. 
Such distributions containing the precise shape information can be 
averaged for representative shapes of an ensemble of particles and 
can be used to calculate the ensemble polarizability
\begin{equation}
\alpha_{av} =  V \int_0^1 P(L) \alpha(L) dL
\end{equation}
with $\alpha$(L) according to Eq. (4). This "distribution of form factors (DFF)" 
model has been proven to be able to successfully describe experimental IR spectra 
for different grain shapes of submicron-sized particles and to predict the 
influence of an embedding medium correctly \cite{mutschke09}. P(L) distributions 
- DFFs - can be derived by fitting experimental spectra and typical DFFs for 
irregular and roundish grains have been extracted (see also Sec.~4.3.). 
Compared to other models, the use of these DFFs allows a fast computation 
of quite realistic band profiles for real particulates. 

Here we should stress that all of the previous considerations are only 
valid in the quasi-static case and for isolated particles. Another 
shortcoming of this and all similar models is that for crystallographic 
anisotropic materials, the polarizability has to be calculated 
with the different sets of optical constants separately. This 
neglects possible interaction of dipoles with different orientations 
and can lead to errors in some cases \cite{mutschke09}.

In dense regions of interstellar matter (dense molecular cloud
cores, circumstellar disks and envelopes), the submicron-sized
grains can coagulate and finally form larger particles \cite{beckwith00,natta07}. 
For a description of the interaction of
electromagnetic radiation with fluffy aggregates composed of
individual particles two distinct approaches are possible:

\begin{enumerate}
\item ``Deterministic'' approach: The frequency-domain Maxwell equation
is solved for an individual cluster. The resulting cross sections are
calculated for many clusters and averaged over the ensemble and the
orientation of the clusters. An advantage of this approach is that for special
systems (e.g., clusters of spheres), exact solutions of the problem do exist.
For computational reasons, however, these methods are limited to either
comparatively small clusters or moderately absorbing systems. For larger
clusters of highly absorbing matter (as in the case of graphitic carbonaceous
grains and metal clusters), one still has to rely on approximations.
Examples of this kind of approach are the discrete dipole and multipole
approximations (DDA/DMA) and the extended Mie theory for multisphere aggregates.

\item ``Statistical'' approach: The equations are formulated in terms of
statistically relevant quantities only (e.g., average radial density function
of the clusters, density correlation function) without any explicit treatment
of the properties of individual particles. Whereas a given cluster generally
does not have any symmetry, statistical averages show rotational invariance,
unless alignment mechanisms (magnetic fields, winds) break this symmetry.
The advantage of this approach is that only the necessary information (ensemble and
orientation averaged quantities) enter
the calculations. Examples of this approach are the different
effective medium theories, the strong permittivity fluctuation theory, and the 
``Distribution of Hollow Spheres'' model \cite{min05}. 
The above-mentioned DFF model belongs also to this class. 

\end{enumerate}

Benchmark results for these methods and astronomically relevant
materials have been published by Stognienko \textit{et al.} \cite{stognienko95}, 
Michel \textit{et al.} \cite{michel96},  Xu \& Gustafson \cite{xugu99}, and 
Voshchinnikov \textit{et al.} \cite{voshch00}. 
The interested reader should consult these papers for more
detailed information. The properties of porous and fractal cosmic
grains have recently found renewed interest and were calculated by 
various authors \cite{voshch05,shen08,shen09,min08}.

The 10 $\mu$m silicate feature has been frequently used to trace
the particle size in protoplanetary disks. Infrared spectroscopy
of disks have shown the presence of flat-top profiles, indicative
of the presence of micron-sized particles which are larger than
the typical particle sizes in the diffuse interstellar medium
(e.g., Refs. \nocite{vboekel05,sicilia07}46 and 47). The influence of porosity 
and aggregate structure on this important astronomical dust 
feature has been investigated by Min \textit{et al.} \cite{min06b} and 
Voshchinnikov \& Henning \cite{vohe08}.

%%%%%%%%%%%%%%%%%%%%%%%%%%%%%%%%%%%%%%%%%%%%%%%%%%%%%%%%%%%%%
\section{Developments in spectroscopic experiments}
\label{sect:experiments}

\subsection{Spectroscopic data of dust minerals}

One type of information which may be possible to extract from infrared 
emission spectra of cosmic dusty media is the chemical composition and 
mineralogical structure of the dust minerals at the present stage of 
evolution of the respective object. In a great effort to provide 
experimental dust spectra for comparison with the astronomical infrared 
data, the spectroscopic properties of the (likely) major dust minerals 
have been investigated in different laboratories and have been made 
available to the astronomical community. For example, the infrared 
spectra of crystalline silicates with different iron to magnesium ratios 
have been studied in great detail in the last decade, especially by the 
group of C. Koike at Kyoto Pharmaceutical University 
(e.g., Refs. \nocite{Chihara02,Koike03}50 and 51) by manufacturing and measuring synthetic 
crystals of variable composition. 

The technique applied for the infrared spectroscopy is usually 
transmission measurements of powders embedded into solid matrices 
(pellets made of KBr, CsI or polyethylene depending on the wavelength 
range of the spectroscopic measurement). In order to reach the Rayleigh 
case (size smaller than wavelength), the particles are mechanically 
ground and size-separated by sedimentation. The fine-grained particles 
are dispersed among the matrix material as homogeneously as possible 
at a certain mass ratio. The pellets then contain the particles at a 
known column density $\sigma = m/A$, with A being the pellet cross 
sectional area. The advantages of this technique are easy preparation, 
small sample consumption and quantitative determination of mass absorption 
coefficients $\kappa = -ln(T)/\sigma$ from the spectral transmission 
of the pellet. Due to Kirchhoffs law, absorption coefficients can be 
directly used as emissivity of dust grains in thermal emission spectra. 

\begin{figure}
\centering
\includegraphics[height=7cm]{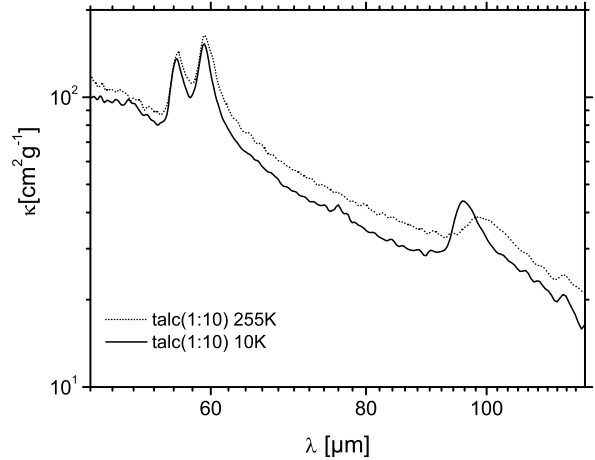}
\caption{Mass absorption coefficient of talc (Mg$_3$[Si$_4$O$_{10}|$(OH)$_{2}$]) 
at far-infrared wavelengths for two different temperatures. The ratio 1:10 
(by mass) refers to the dilution of the talc powder by polyethylene in the 
pellet --- after Ref. 55.} \label{fig-3}
\end{figure}

The same technique has been employed to study the variations of band 
positions and band profiles with temperature, especially for low 
temperatures down to a few Kelvin. Because of the anharmonicity of 
the vibrational potentials of the lattice modes causing the infrared 
bands, the bands tend to get narrower and to shift to smaller wavelengths 
as the vibrational levels are de-populated with lowering T. Such studies 
have been performed e.g., by the groups in London, Kyoto, St. Louis and 
Jena for olivine particles \cite{Bowey01,Chihara01,Koike06} and for 
phyllosilicates (see Fig.~\ref{fig-3}) \cite{Koike90,Hofmeister06,mutschke08}. 
It has been found that the sharpness and wavelength position of certain 
silicate far-infrared bands might be actually good indicators of the dust 
temperature. 

The drawback of the powder transmission spectroscopy in pellets, generally, 
is the possible influence of the polarisation of the embedding medium on 
the spectra. It can be easily shown for instance by Mie calculations that 
such an influence exists (see Sec.~3). In the general case of nonspherical 
or even irregular particle shapes, however, it is difficult to correct for 
this influence, although it is possible if the material's optical constants 
are known (see also below) \cite{papoular98,clement03}. 

\subsection{Optical constants}

Most astrophysicists modeling thermal dust emission spectra rely on 
calculated rather than experimentally measured spectra, 
because of the flexibility of such calculations if parameters such 
as grain sizes have to be varied. The material properties enter into 
these calculations usually in the form of the so-called ``optical 
constants''. This term denotes the response function which characterizes 
the reaction of a material to an external (time-variable) electromagnetic 
field: the complex refractive index $m=n+ik$. If magnetic material properties 
can be neglected, $m$ is the square root of the (also complex) dielectric 
function $\varepsilon$ which alternatively can be used as optical constants. 

\begin{figure}
\centering
\includegraphics[height=7cm]{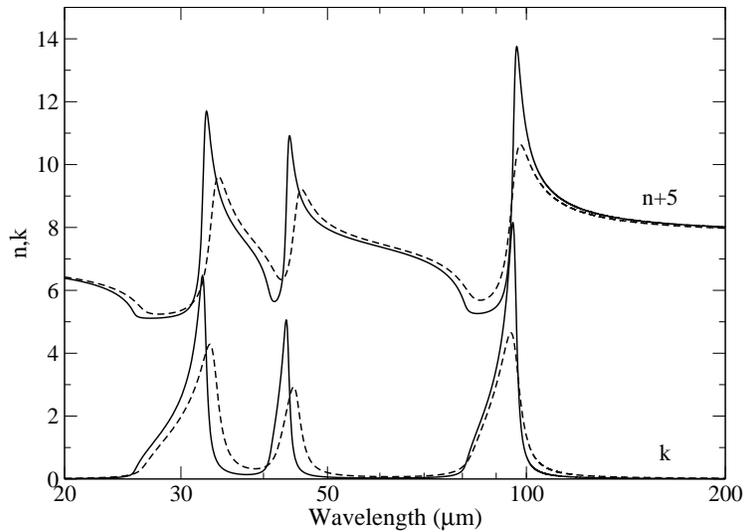}
\caption{Far-infrared optical constants of calcite at temperatures of 
10~K (solid lines) and 300~K (dashed lines) --- after Ref. \nocite{posch07}61.} 
\label{fig-4}
\end{figure}

It has to be noted that these quantities are macroscopic in character, 
therefore they loose their meaning for small clusters and molecules. 
In the transition region from solids to molecules the introduction of 
size-dependent optical material properties may be a reasonable way 
to include e.g., the confinement of charge carriers to the limited 
particle volume \cite{kreibig95}. 

The term optical constants is a bit misleading since they are 
strongly frequency-dependent. The dispersion with frequency is 
determined by resonances of the electronic system, of the ionic lattice 
and, at very low frequencies, by relaxation of permanent dipoles. 
In these frequency regions absorption becomes strong (high $k$) and 
$n$ shows ``anomalous dispersion'', i.e. decreases with frequency. 
In many cases, this behaviour can be described by Lorentzian oscillators. 

Frequency-dependent optical ``constants'' of solid materials can be found 
in a number of databases which are available either in the form of books or 
electronic media. The most important database in book-form is the 
``Handbook of Optical Constants of Solids'' edited by E.D. Palik that 
currently consists of three volumes which appeared in 1985, 1991 
and 1998 \cite{palikI,palikII,palikIII}. These books are recommended 
since they comprise detailed discussions of the origin and errors of 
each dataset which may prevent the user from uncritical application of 
these data. 

The Jena-St.Petersburg-Heidelberg database of optical constants 
\cite{henning99} which is especially dedicated to cosmic dust 
applications can be found via the web pages http://www.astro.uni-jena.de 
and http://www.mpia-hd.mpg.de/HJPDOC.

As mentioned already, nearly all of these optical constants have been 
measured either on bulk samples or on thin films. The reason for 
that is that the planar geometry of films or coatings is much easier 
to describe in deriving optical constants from a measured spectrum 
than are irregularly shaped particles. However, even in the 
case of ideal bulk measurements, the determination of 
optical constants over a wide frequency or wavelength range 
is not a simple task. Since the material absorption in different 
spectral regions usually differs by many orders of magnitude, 
for the determination of $k$ either transmission measurements on 
samples of very different thicknesses (from centimeter down to 
submicron scales) or transmission and reflection measurements 
have to be combined \cite{dorschner95}. Many crystals show 
an anisotropy in their optical constants. In these cases, 
measurements with polarized light along the different axes of 
the crystal have to be carried out which require careful orientation 
of the crystal and alignment of the polarizers.

In regions of strong absorption, as mainly considered here, usually 
reflection measurements have to be used and either the Kramers-Kronig 
relations (see Ref. 24) or a parametrization of the optical 
constants as a sum of Lorentzian oscillators have to be applied in 
order to obtain both the real and imaginary parts of the optical constants.  
As already mentioned above, the optical constants of cosmic dust analogs 
may strongly depend on the temperature of the material, so that 
temperature-dependent data are desired. 

Despite of the difficulties inherent to the preparation of samples and 
the measurement and data evaluation procedures in obtaining such data, 
some progress has also been made in the measurements of the optical 
constants for dust materials within the last years. For instance, optical 
constants of amorphous magnesium silicates with a variable magnesium 
content have been measured in a wide wavelength range \cite{jaeger03}. 
For crystalline silicates, the optical constants of minerals of the 
olivine series (composition (Mg,Fe)$_2$SiO$_4$) have been re-measured 
for different magnesium-to-iron ratios (e.g., Refs. \nocite{fabian01,Sogawa06}68 and 69) 
and even at temperatures down to 50~K (for Mg$_2$SiO$_4$, \cite{Suto06}). 
More temperature-dependent optical constants have been derived for 
carbonates \cite{posch07}, including the far-infrared wavelength range 
(see Fig.~\ref{fig-4}). 

\subsection{Aerosol measurements and grain morphology}

The influence of the grain morphology and of the embedding into 
matrices has always been a concern in the use of experimental spectra 
for comparison with observed ones. As demonstrated in Sec.~3, 
the absorption band profiles of solid particles depend on the 
grain shape and on the electromagnetic polarisability 
(the refractive index) of the ambient medium. 

When comparing spectra of crystalline silicates measured in different 
laboratories, it has been realized that band profiles and peak positions 
showed systematic differences, which were supposedly related to the 
morphological properties of the particulates such as the grain shape 
(influenced by the powder grinding technology) and perhaps the (unknown) 
state of agglomeration inside the pellets \cite{molster02,Koike06}. 
On the other hand, spectra calculated from the material's optical 
constants assuming a spherical grain shape failed completely to reproduce 
both observed and measured spectra. Simple statistical approaches to more 
realistic grain shape models such as the Continuous Distribution of 
Ellipsoids (see Sec.~3) gave much better, but still not sufficient, 
results \cite{fabian01}. 

\begin{figure}
\centering
\includegraphics[height=9cm]{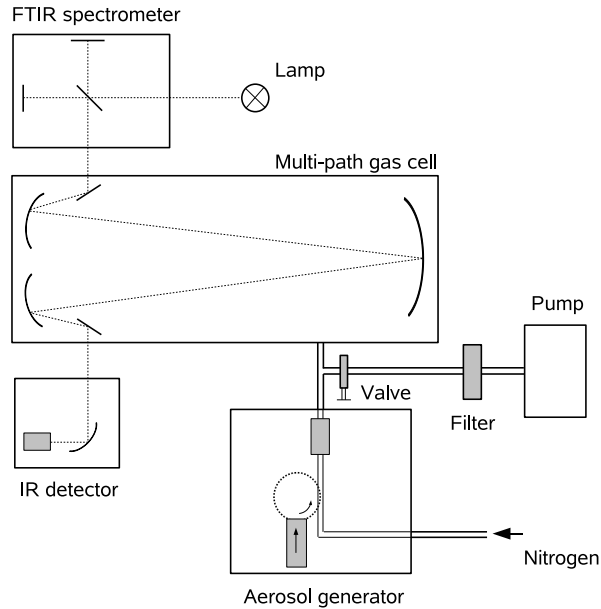}
\caption{Schematic structure of the aerosol spectroscopy setup.} \label{fig-5}
\end{figure}

\begin{figure}
\centering
    \includegraphics[height=7cm]{henning_mutschke6a.eps}
    \includegraphics[width=7cm]{henning_mutschke6b.eps}
    \includegraphics[width=6cm]{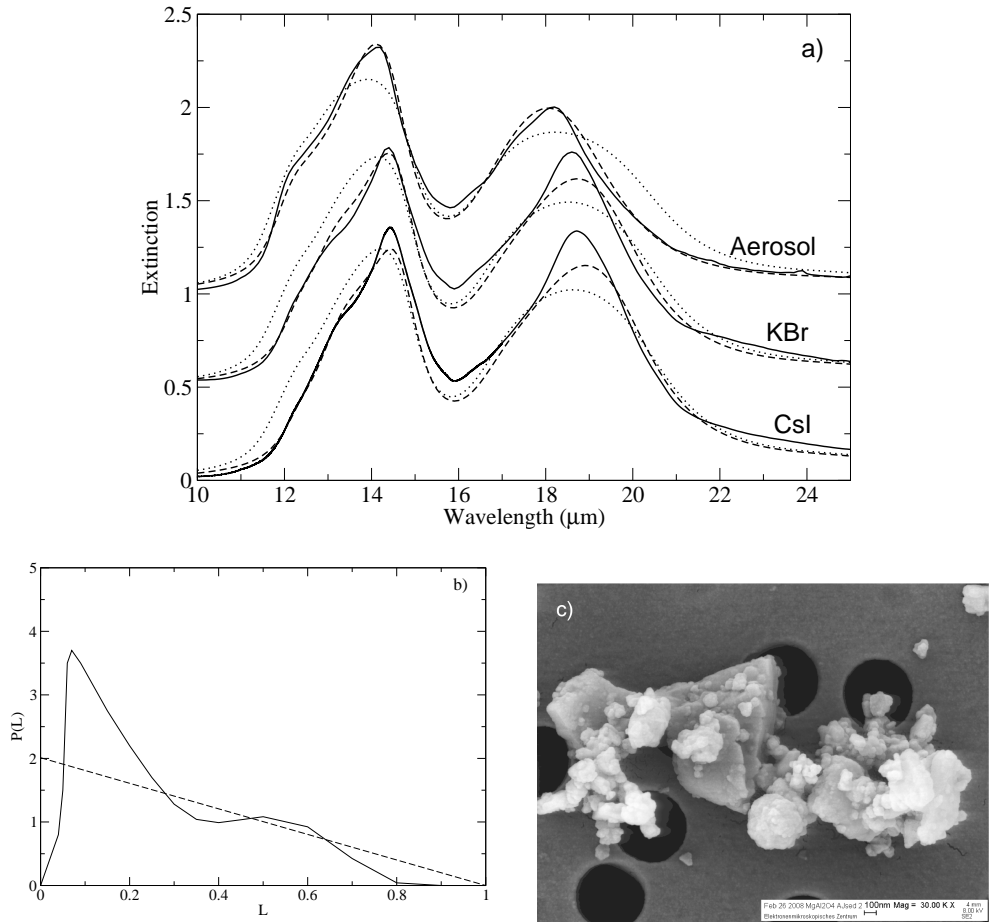}
\caption{(a) Absorption spectra of irregular-shaped spinel particles 
measured in different media (nitrogen gas, KBr and CSi pellets 
- solid lines) compared to simulated spectra using the CDE 
(dotted line) and DFF (dashed line) models. (b) The shape factor 
distribution used for the DFF calculations (solid line) compared 
to that of the CDE (P(L) = 2(1-L), dashed line). (c) Electron micrograph 
of the particles. The black filter holes have a diameter of 0.5~$\mu$m.}
\label{fig-6}
\end{figure}

From the experimental side, an important step forward was made a few 
years ago, when the aerosol spectroscopy technique was introduced into 
the measurement of experimental comparison spectra for astronomical puposes 
\cite{Tamanai06a}. In contrast to the pellet method, this technique 
proved to be suitable not only for avoiding any influence of an embedding 
medium onto the spectra, but also for a detailed study of morphological 
effects on infrared spectra of dust samples \cite{Tamanai06b}. 
The reason for this is that a direct electron microscopic investigation 
of the aerosol particles is routinely possible after the spectroscopic 
measurement. 

The experimental setup for aerosol infrared spectroscopy as used at the 
astrophysical laboratory of the AIU Jena consists of a White-type multi-path 
gas cell connected to a Bruker 113v FTIR spectrometer and equipped with 
an ``aerosol generator'' for dispersing fine powders in nitrogen gas and 
filling the gas cell with this aerosol (see Fig.~\ref{fig-5}). The aerosol 
generator makes use of a Palas RBG 1000 rotating-brush particle disperser 
and is also equipped with an impactor-type grain-size discriminator in 
order to allow only small particles to reach the gas cell. The aerosol
 formed of these submicron-sized particles in a gas at about normal 
pressure is stable for several minutes, sufficient for FTIR measurements. 
After the measurement, a part of the particles are filtered from the 
aerosol for scanning electron microscopic (SEM) inspection. The drawbacks 
of this technique are that the column density of the particles is not 
precisely known, making a quantitative measurement of the absorption 
coefficient difficult, the large sample consumption, and the limitation 
to a certain particle size determined by the sedimentation speed. 

By comparing the aerosol-measured infrared spectra with pellet spectra, 
the influence of the embedding medium on the band profiles has been made 
clear for a large number of oxide \cite{Tamanai09a} and silicate 
\cite{Tamanai09b} mineral ``dust samples''. The internet database 
making such aerosol spectra publically available (http://elbe.astro.uni-jena.de) 
contains also spectra of particles embedded in cesium iodide (CsI) 
pellets for comparison. The latter are usually characterized by infrared 
bands peaking at longer wavelengths. This does not mean that the bands are 
just shifted, but that surface modes closer to the transverse optical 
lattice resonance get a higher weight in the band profiles (see below). 

In the course of these aerosol measurements, great attention has been 
given to morphological effects in the spectra and it has been confirmed 
that grain shapes play a big role for the band profiles. For several 
compounds it has been possible to measure powders with different 
grain shapes, for instance more roundish ones vs. more irregular, 
sharp-edged shapes. The respective infrared bands were found to be 
characteristically different, for the roundish grains often revealing 
a short-wavelength peak and a long-wavelength shoulder, whereas 
irregular grains produced long- or intermediate wavelength single 
peaks \cite{mutschke09,Tamanai09a}. 

As mentioned in Sec.~3, the DFF model can successfully be used to 
reproduce these band profiles. Figure~\ref{fig-6} demonstrates this 
for irregularly shaped spinel (MgAl$_2$O$_4$) particles measured 
in various environments (aerosol, pellets). The form factor 
distribution used in these computations is also shown. It has to 
be noted again that this model, as many others, is only exact in 
the case of particles composed of an isotropic material. 

\section{Conclusions}
The measurements and the theoretical calculations of optical properties of cosmic grains play an
important role in a wide range of astrophysical applications. The collection of relevant data
for well-characterized materials over a broad wavelength range has made great progress over the 
last two decades. Here we want to stress that the characterization of the physical and chemical 
structure of the materials is an important part of the study of cosmic dust analogs. An 
interesting direction of future research will be the comparison of the properties of 
cosmic dust grains collected by interplanetary space missions such as the Stardust mission 
(e.g., Ref. \nocite{bradley10}76) with synthetic materials produced in the laboratory. 

The combination of condensation experiments carried out under conditions similar to cosmic environments
and advanced spectroscopic methods is certainly the future of solid-state astrophysics. Mass
spectroscopy has to be used to characterize the transition from atoms and molecules to small
clusters and solids. In addition, the change of material properties
under the bombardment of cosmic rays and irradiation with intense X-ray and UV photons
has to be better understood. Another important factor can be thermal annealing cycles
of the grains.

The application of laboratory data for the interpretation of astronomically 
measured fluxes relies on the knowledge of the density and temperature structure of the considered 
dusty regions. 
Radiative transfer calculations have to be carried out for deeply embedded objects
such as protostars and active galactic nuclei. In the case of very small particles the
effects of quantum heating have to be taken into account. 
Even for the interpretation of spectra of optically thin configurations, the dust temperature 
and density distributions have to be known.

%%%%%%%%%%%%%%%%%%%%%%%%%%%%%%%%%%%%%%%%%%%%%%%%%%%%%%%%%%%%%
\acknowledgments
We gratefully acknowledge the collaboration with our colleagues Dr. Akemi Tamanai, Gabriele Born, 
and Walter Teuschel (all AIU Jena) on the laboratory experiments and Dr. Michiel Min (Astronomical 
Institute Utrecht) on the simulations of dust spectra. Our research has been supported by the 
Deutsche Forschungsgemeinschaft with grants MU 1164/5,6.

%%%%%%%%%%%%%%%%%%%%%%%%%%%%%%%%%%%%%%%%%%%%%%%%%%%%%%%%%%%%%
%%%%% References %%%%%
\newcommand{\aap}{Astron. Astrophys.}
\newcommand{\AAp}{\aap}
\newcommand{\aaps}{Astron. Astrophys. Suppl.}
\newcommand{\AApS}{\aaps}
\newcommand{\apj}{Astrophys. J.}
\newcommand{\ApJ}{\apj}
\newcommand{\apjl}{\apj\ Lett.}
\newcommand{\ApJL}{\apjl}
\newcommand{\apjs}{\apj\ Suppl.}
\newcommand{\ApJS}{\apjs}
\newcommand{\aj}{The Astronomical Journal}
\newcommand{\AJ}{\aj}
\newcommand{\jgr}{Journal of Geographic Research}
\newcommand{\JGR}{\jgr}
\newcommand{\mnras}{Mon. Not. R. Astron. Soc.}
\newcommand{\MNRAS}{\mnras}
\newcommand{\AdvSR}{\apspr}
\newcommand{\planss}{Planetary and Space Science}
\newcommand{\PSS}{\planss}
\newcommand{\pasp}{Publications of the Astronomical Society of the Pacific}
\newcommand{\PASP}{\pasp}
\newcommand{\nat}{Nature}
\newcommand{\Nat}{Nature}
\newcommand{\araa}{Annu. Rev. Astron. Astrophys.}
\newcommand{\ARAA}{\araa}
\newcommand{\areps}{Annual Review of Earth and Planetary Science}
\newcommand{\AREPS}{\areps}
\newcommand{\aapr}{Astron. Astrophys. Rev.}
\newcommand{\AApR}{\aapr}
\newcommand{\qjras}{Quarterly Journal of the Royal Astronomical Society}
\newcommand{\ao}{Applied Optics}
\newcommand{\pasj}{Proc. Astron. Soc. Jpn.}
%\bibliography{references}   %>>>> bibliography data in report.bib
%\bibliographystyle{spiejour}   %>>>> makes bibtex use spiejour.bst

%%%%%%%%%%%%%%%%%%%%%%%%%%%%%%%%%%%%%%%%%%%%%%%%%%%%%%%%%%%%%
%%%%% Biographies of authors %%%%%

%\vspace{2ex}\noindent{\bf John P. Doe} is an assistant professor at the University of Ohio. He received his BS and MS degrees in physics from the University of Washington in 1985 and 1987, respectively, and his PhD degree in optics from the Rochester Institute of Technology in 1991.  He is the author of more than 50 journal papers and has written three book chapters. His current research interests include optical interconnects, holography, and optoelectronic systems. He is a member of SPIE.

%\vspace{2ex}\noindent{\bf Jane C. Smith}: biography not available.

\end{document}